\documentclass[pra,aps,reprint,superscriptaddress,a4paper]{revtex4-1}
\usepackage{amsmath}
\usepackage{braket}
\usepackage{graphicx}
\usepackage{verbatim}
\usepackage{color}
\usepackage{hyperref}
\begin{document}
\title{
	Finite-temperature properties of interacting bosons on a two-leg flux ladder
}
\author{
	Maximilian Buser
}
\affiliation{
	Department of Physics,
	Arnold Sommerfeld Center for Theoretical Physics (ASC),
	Munich Center for Quantum Science and Technology (MCQST),
	Fakult\"{a}t f\"{u}r Physik, Ludwig-Maximilians-Universit\"{a}t M\"{u}nchen,
	80333 M\"{u}nchen, Germany
}
\author{Fabian Heidrich-Meisner}
\email{Corresponding author: heidrich-meisner@uni-goettingen.de}
\affiliation{
	Institute for Theoretical Physics, Georg-August-Universit\"at G\"ottingen, 37077 G\"ottingen, Germany
}
\author{Ulrich Schollw\"ock}
\affiliation{
	Department of Physics,
	Arnold Sommerfeld Center for Theoretical Physics (ASC),
	Munich Center for Quantum Science and Technology (MCQST),
	Fakult\"{a}t f\"{u}r Physik, Ludwig-Maximilians-Universit\"{a}t M\"{u}nchen,
	80333 M\"{u}nchen, Germany
}
\date{\today}
\begin{abstract}
Quasi-one-dimensional lattice systems such as flux ladders with artificial gauge fields host rich quantum-phase diagrams that have attracted great interest. 
However, so far, most of the work on these systems has concentrated on zero-temperature phases while the corresponding finite-temperature regime remains largely unexplored.
The question if and up to which temperature characteristic features of the zero-temperature phases persist is relevant in experimental realizations.
We investigate a two-leg ladder lattice in a uniform magnetic field and concentrate our study on chiral edge currents and momentum-distribution functions, which are key observables in ultracold quantum-gas experiments. 
These quantities are computed for hard-core bosons as well as noninteracting bosons and spinless fermions at zero and finite temperatures.
We employ a matrix-product-state based purification approach for the simulation of strongly interacting bosons at finite temperatures and analyze finite-size effects.
Our main results concern the vortex-fluid-to-Meissner crossover of strongly interacting bosons.
We demonstrate that signatures of the vortex-fluid phase can still be detected at elevated temperatures from characteristic finite-momentum maxima in the momentum-distribution functions, while the vortex-fluid phase leaves weaker fingerprints in the local rung currents and the chiral edge current.
In order to determine the range of temperatures over which these signatures can be observed, we introduce a suitable measure for the contrast of these maxima.
The  results are condensed into a finite-temperature crossover diagram for hard-core bosons.
\end{abstract}
\maketitle
%
%
%
%
\section{Introduction}
The recent realization of artificial gauge fields in optical-lattice experiments with ultracold neutral Bose gases has attracted great interest~\cite{goldman2016topological,goldman_2014,galitski_2013,dalibard_2011,cooper2018topological,aidelsburger2017artificial}.
State-of-the-art implementations are based on the utilization of so-called synthetic dimensions created by Raman coupling of internal atomic states~\cite{lin_2009_nature,lin_2009_prl}, laser-assisted tunneling \cite{aidelsburger_2011,aidelsburger_2013,ketterle_2013}, or 
periodically modulated optical lattices \cite{sengstock_2012,esslinger_2014,Flaschner2016,Asteria2018}.
While experimental research mainly focused on noninteracting particles, recent advances encourage the prospect of future experiments with strongly interacting bosons~\cite{Kennedy2015,Lohse2016,greiner_2017}.
%

\begin{figure}
        \includegraphics[]{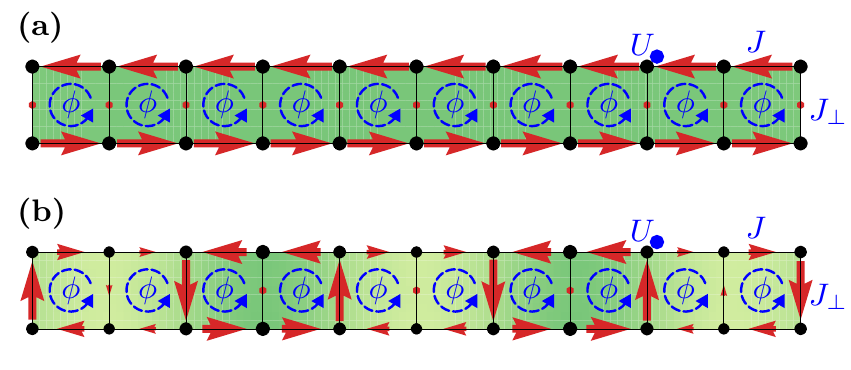}
        \caption{
                {\it Meissner and vortex-fluid phase.}  Ground-state particle current patterns of noninteracting bosons for
                (a) Meissner phase and
                (b) vortex-fluid  phase.
                Red arrows indicate the direction and, by their length, the strength of the local currents.
                The size of the dots and the background shading represent the local particle density.
                The Hamiltonian parameters $\phi$, $J_\perp$, $J$ and $U$ are introduced in Eq.~(\ref{ham_2legladder_runggauge}).
                A comprehensive analysis of the Meissner and vortex-fluid phases can be found in,  e.g., Refs.~\cite{giamarchi_2001,petrescu_lehur_2013,paredes_2014,georges_2014,piraud_2015}.
        }
        \label{sketch_current_patterns}
\end{figure}

%
In this context, quasi-one-dimensional ladderlike systems (see Fig.~\ref{sketch_current_patterns} for a sketch) are particularly timely.
The corresponding geometries can be isolated in superlattice potentials~\cite{atala_2014} and are naturally realized in setups employing a synthetic dimension~\cite{stuhl_2015,fallani_2015,kolkowitz2017spin,Wall2016,an2017direct,kang2018realization}.
Moreover, ladderlike lattices host rich physics and allow  for a direct comparison between theoretical predictions for interacting systems and experimental results.
While originally advocated as a suitable system to measure edge states in experiments \cite{paredes_2014,atala_2014}, amazingly, it has also been suggested that  Chern numbers can be mapped out experimentally in multileg flux ladders with as few as five legs and open boundary conditions~\cite{tarruell_scipost2017}, which was studied in a recent experiment \cite{genkina2018imaging}.
%

%
Analytical as well as numerical studies, with a large fraction of the latter work based on the density-matrix renormalization-group method~\cite{White1992,schollwock2011density}, have provided extensive theoretical results regarding the ground-state properties of interacting quantum gases on such ladderlike systems~\cite{giamarchi_2001,carr_2006,roux_2007,dhar_2012,dhar_2013,petrescu_lehur_2013,georges_2014,grusdt2014,wei_mueller_2014,paredes_2014,petrescu_lehur_2015,di_dio_2015,cornfeld_2015,piraud_2015,greschner_2015_prl,Bilitewski2016,greschner_2016,petrescu_piraud_lehur_2017,Greschner2017,uchino_2015,uchino_2016,orignac_2017,huegel17,citro_2018,strinati_2018,greschner_giamarchi_2018,coira2018finite,greschner_giamarchi_2018}.
The presence of gauge fields clearly enriches the corresponding phase diagrams, which by now have been mapped out in large part.
For instance, these phase diagrams host superfluid and Mott-insulating vortex-liquid and Meissner phases~\cite{giamarchi_2001,petrescu_lehur_2013,piraud_2015} and vortex lattices~\cite{giamarchi_2001,greschner_2015_prl,greschner_2016}, as well as charge-density-wave~\cite{greschner_2016,Greschner2017} and biased-ladder phases~\cite{wei_mueller_2014}.
Typical particle-current patterns found in the Meissner and vortex-fluid phases are shown in Figs.~\ref{sketch_current_patterns}(a) and \ref{sketch_current_patterns}(b), respectively.
Furthermore, the possible existence of fractional quantum Hall-like states has attracted great interest~\cite{grusdt2014,petrescu_lehur_2015,cornfeld_2015,petrescu_piraud_lehur_2017,strinati_2017}.
Recently, results regarding the steady state of a two-leg ladder system subjected to a dissipative environment have been put forward in Ref.~\cite{halati_2017}, and the ground-state phase diagram of a related but more complicated two-leg Haldane model has been mapped out in Ref.~\cite{greschner_2018}.
We note that currents in mesoscopic ring-shaped flux ladder systems, which have been experimentally realized in Ref.~\cite{amico14}, have been studied in Ref.~\cite{haug18}.
However, most of these studies concentrated on ground states, while experiments are naturally at nonzero energy densities.
%

%
In this paper, we investigate a two-leg ladder in a uniform Abelian gauge field at finite temperatures $T$.
Especially, we concentrate on the most prominent transition, the vortex-fluid-to-Meissner crossover of hard-core bosons as well as of noninteracting bosons and a related transition for spinless fermions at $T>0$~\cite{giamarchi_2001,carr_2006,roux_2007,petrescu_lehur_2013,paredes_2014,piraud_2015,di_dio_2015}.
We compute chiral edge currents and momentum-distribution functions at zero and finite temperatures and determine the range of temperatures in which clear signatures of the vortex-fluid phase can be detected.
For this purpose, we employ a matrix-product-state based purification approach~\cite{schollwock2011density,Verstraete_2004}, which we implement for both canonical setups with fixed numbers of particles as well as grand-canonical setups with a fixed average number of particles.
We note that aspects of finite-temperature physics in ladder systems have previously been investigated in Refs.~\cite{greschner_2015_prl,strinati_2017,coira2018finite,citro_2018}.
%

%
We first revisit the case of noninteracting bosons and fermions \cite{carr_2006,paredes_2014,georges_2014} and discuss the corresponding vortex(-fluid)-to-Meissner transition.
Moreover, in these limits, we can compute results at finite temperatures for very large systems and, therefore, we are able to obtain estimates for the typical finite-size effects. 
Several different quasimomentum distribution functions are introduced and we argue that these are better suited to detect the vortex fluid at  $T>0$ than the chiral edge current or local rung currents. 
We introduce a quantitative measure that detects the presence of finite-momentum peaks in a suitably chosen momentum-distribution function.
Moreover, we compare chiral edge currents observed in canonical and grand-canonical setups.
Our main results are for hard-core bosons, for which we present a finite-temperature crossover state diagram for model parameters realized in \cite{atala_2014}.
Finally, we point out other interesting questions for future finite-temperature studies concerning, e.g., the melting of vortex lattices or of the biased-ladder phase at $T>0$ \cite{greschner_2015_prl,greschner_2016} in the  strongly interacting, low-density regime.
%

%
In the following, we first define  the Hamiltonian of the ladder system and introduce observables of interest  in Sec.~\ref{sec_model}.
There, we also recap key aspects of the noninteracting model.
The purification approach used for the calculation of thermal states of strongly interacting bosons is introduced in Sec.~\ref{sec_purification}.
Our main results concerning the vortex-fluid-to-Meissner crossover at finite temperatures are presented in Sec.~\ref{main_results}.
There, we discuss chiral edge currents and  momentum-distribution functions of hard-core bosons and noninteracting spinless fermions, as well as noninteracting bosons at zero and finite temperatures.
Finally, we summarize our work in Sec.~\ref{summary}.
%
%
%
%
\section{Model and observables of interest}
\label{sec_model}
\subsection{Two-leg ladder Hamiltonian}
This paper mainly focuses on a Bose-Hubbard Hamiltonian defined on a two-leg ladder geometry.
Using annihilation (and creation) operators $b^{(\dagger)}_{l,r}$ referring to bosons or spinless fermions located on site $\left(l,r\right)$, the Hamiltonian reads
\begin{align}
H=
&-J\left(\sum_{l=0}^{1}\sum_{r=0}^{L-2}b_{l,r}^\dagger b_{l,r+1}+\text{h.c.}\right)\nonumber\\
&-J_\perp\left(\sum_{r=0}^{L-1}e^{-ir\phi}b_{0,r}^\dagger b_{1,r}+\text{h.c.}\right)\nonumber\\
&+\frac{U}{2}\sum_{l=0}^{1}\sum_{r=0}^{L-1} b_{l,r}^\dagger b_{l,r}\left(b_{l,r}^\dagger b_{l,r}-1\right).\label{ham_2legladder_runggauge}
\end{align}
Here, the indices $l=0,1$ and $r=0,\dots,L-1$ label the legs and rungs, respectively,  of a two-leg ladder with  $L$ rungs.
The first term in Eq.~(\ref{ham_2legladder_runggauge}) represents nearest-neighbor particle hopping along the legs of the ladder.
We use the corresponding hopping amplitude $J$ as our unit of energy (we set $\hbar=1$ and $k_B=1$).
The second term accounts for interleg hopping along the rungs of the ladder. 
It is designed in such a way that the state of a particle encircling a single plaquette of the ladder gains a phase factor $e^{\pm i\phi}$, with the sign depending on the direction of the circulation.
The last term in Eq.~(\ref{ham_2legladder_runggauge}) accounts for repulsive interactions between two or more bosons located at the same site (for spinless fermions, this term is absent). 
In this paper, we concentrate on hard-core bosons with a maximum occupation of one particle per site ($U/J\rightarrow\infty$) and noninteracting particles ($U=0$).
Moreover, in the examples presented throughout this paper, we consider a value of $\phi=\pi/2$, which  corresponds to the experimental setup of Ref.~\cite{atala_2014}.
Thereby, the considered Hamiltonian exhibits a translational symmetry with a unit cell containing four plaquettes of the ladder in the (rung) gauge chosen in Eq.~\eqref{ham_2legladder_runggauge}.
\subsection{Noninteracting model}
\label{sec:nonint}
Next, we recap some of the key aspects of the noninteracting ($U = 0$) flux-ladder model with periodic boundary conditions. 
While these are well-known~\cite{georges_2014,paredes_2014}, their discussion is given here for the sake of completeness and will guide the analysis of finite-temperature properties of noninteracting fermions and bosons in Sec.~\ref{main_results}.
The corresponding Hamiltonian is denoted by $H_F$.
%

\begin{figure}
	\includegraphics[]{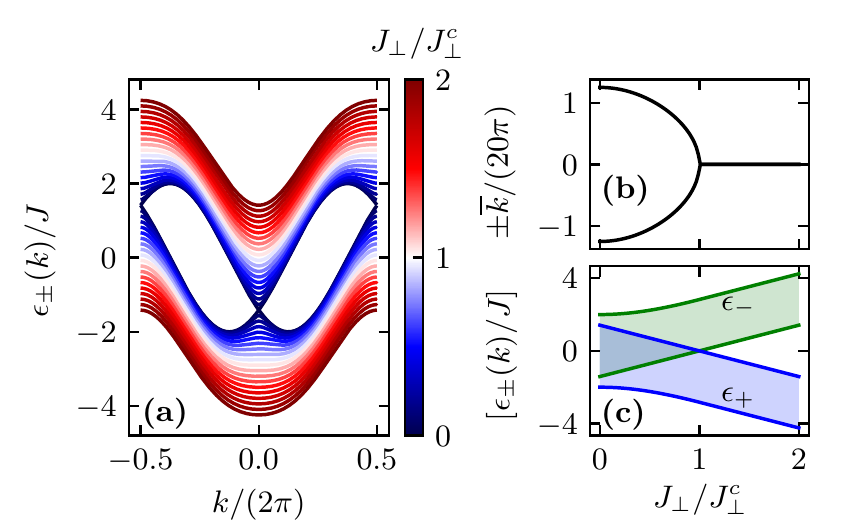}
	\caption{
		{\it Band structure of the noninteracting model with periodic boundary conditions}.
		(a)	The two bands $\epsilon_{\pm}$ for different values of $J_\perp/J_\perp^c$.
		In the Meissner phase ($J_\perp>J_\perp^c$), the lower band exhibits a single minimum at zero momentum.
		In the vortex-fluid phase ($J_\perp<J_\perp^c$), the lower band exhibits two degenerate minima at finite momenta.
		The lower and upper band are separated by a finite-energy gap for nonvanishing values of $J_\perp$.
		(b) The characteristic momenta $\pm\overline{k}$ of the ground states clearly indicate the vortex-fluid-to-Meissner transition.
		(c) The shaded regions represent the ranges $\left[\epsilon_{\pm}(k)\right]$ of the lower and upper band.
		A comprehensive analysis of the noninteracting problem can be found in Refs.~\cite{georges_2014,paredes_2014}.
	}
	\label{nonint}
\end{figure}

%
To diagonalize $H_F$, we first rewrite it in terms of leg-gauge operators denoted by an overhead tilde and defined by $\tilde{b}_{0,r} = e^{ir\frac{\phi}{2}}b_{0,r}$ and $\tilde{b}_{1,r} = e^{-ir\frac{\phi}{2}}b_{1,r}$, which moves the complex hopping amplitudes from the rungs to the legs of the system.
Second, leg-gauge momentum operators $\tilde{b}^{(\dagger)}_{l,k_m}$ are introduced by means of Fourier transforming along the legs of the ladder,
\begin{align}
\tilde{b}_{l,k_m} &= \frac{1}{\sqrt{L}}\sum_{r=0}^{L-1}e^{i k_m r}\tilde{b}_{l,r},\label{leg-gauge_mom_op}
\end{align}
with $m=0,\dots,L-1$ and momenta $k_m=2\pi m/L$.
In terms of leg-gauge momentum operators, the noninteracting Hamiltonian with periodic boundary conditions takes the form
\begin{align}
H_F &= -\sum_{m=0}^{L-1}
\begin{pmatrix}
\tilde{b}_{0,k_m}^\dagger& 
\tilde{b}_{1,k_m}^\dagger
\end{pmatrix}
h(k_m)
\begin{pmatrix}
\tilde{b}_{0,k_m}\\ 
\tilde{b}_{1,k_m}
\end{pmatrix}
\intertext{with}
h(k) &=
\begin{pmatrix}
2J\cos\left(\frac{\phi}{2}+k\right)&J_\perp\\
J_\perp&2J\cos\left(\frac{\phi}{2}-k\right)
\end{pmatrix}.
\end{align}
Finally, the operators $d_{\pm,k_m}^{(\dagger)}$ diagonalizing $H_F$ read
\begin{align}
d_{\pm,k_m} = \frac{\left(A_{k_m}\mp B_{k_m}\right)\tilde{b}_{0,k_m}\pm\xi\tilde{b}_{1,k_m}}{\sqrt{\left(A_{k_m}\mp B_{k_m}\right)^2+\xi^2}},
\end{align}
with $\xi=\frac{J_\perp}{2J}$, $B_{k}=\sin(k)\sin(\phi/2)$ and $A_{k}=\sqrt{\xi^2+B_k^2}$.
Thus, $H_F$ takes the form 
\begin{align}
H_F = \sum_{m=0}^{L-1}\sum_{s=+,-} \epsilon_{s}(k_m) d_{s,k_m}^\dagger d_{s,k_m}, 
\end{align}
where the eigenvalues corresponding to the lower and upper band, $\epsilon_{+}(k_m)$ and $\epsilon_{-}(k_m)$, are given by $\epsilon_{\pm}(k_m)=-2J\left(\cos(k)\cos(\phi/2)\pm A_{k_m}\right)$.
For a nonvanishing interleg coupling $J_\perp\ne 0$, the two bands are separated by a finite gap.
Depending on the parameters $\phi$ and $J_\perp$, the lower band has either a single global minimum at momentum $k=0$ or two degenerate minima at $k=\pm\overline{k}$ with
\begin{align}
\overline{k} = 
\arccos \left(
\frac{ \sqrt{  \xi^2 + \sin\left(\phi/2\right)^2 } }
{\tan\left(\phi/2\right)} 
\right).\label{nonint_kmin}
\end{align}
For $k=0$, the system is in the Meissner phase, while for $k=\pm\overline{k}$ it is in the vortex-fluid phase \cite{paredes_2014,georges_2014}. 
The corresponding degenerate single-particle ground states are denoted by $\ket{\pm\overline{k}}$. 
For a given value of $\phi$, the phases are separated by $J_\perp^c=2J\sin\left(\phi/2\right)\tan\left(\phi/2\right)$.
The  two-band structure of the noninteracting ladder model is shown for different values of $J_\perp$ and $\phi=\pi/2$ in Fig.~\ref{nonint}(a).
Figure~\ref{nonint}(b) shows the characteristic momenta $\pm\overline{k}$ of the ground states as a function of $J_\perp$.
%

%
In anticipation of the  discussion of noninteracting fermions, Fig.~\ref{nonint}(c) illustrates the range $\left[\epsilon_{\pm}(k)\right] = \left[\min_{k}\epsilon_{\pm}(k),\max_{k}\epsilon_{\pm}(k)\right]$ of the lower and upper band for $\phi=\pi/2$.
From this, a transition-point (for $\phi=\pi/2$) for the fermionic system  can be  determined from
\begin{align}
\min\limits_{k}\epsilon_{-}(k) &< \max\limits_{k}\epsilon_{+}(k)&&\text{(vortex phase)},\\
\min\limits_{k}\epsilon_{-}(k) &> \max\limits_{k}\epsilon_{+}(k)&&\text{(Meissner)}.\label{band_overlap}
\end{align}
We use the same terminology -- Meissner and vortex phase -- as for bosons since the local current patterns show the same phenomenology:
a chiral edge current with no rung currents in the Meissner phase and finite rung currents in the vortex phase for open boundary conditions
(data not shown; see the discussion of the chiral edge current in Sec.~\ref{sec:jc_fermions}).
We will use these equations to elucidate the vortex-to-Meissner transition of noninteracting fermions at filling one-half.
Note that we define filling $f$ as the number of particles divided by $2L$.
The fermionic transition is further related to a metal (=``vortex phase") to band-insulator (=``Meissner") transition~\cite{carr_2006}.
\subsection{Observables of interest}
\subsubsection{Particle currents}
In the context of the ladder model, local particle currents are of special importance. 
They have been successfully employed as a key measure for  distinguishing between different zero-temperature phases (see \cite{greschner_2016} for an overview).
Moreover, they are experimentally accessible as demonstrated in various cold-atom experiments, see, for instance, Refs.~\cite{atala_2014,stuhl_2015,fallani_2015}.
%

%
The continuity equation for the occupation of lattice sites implies that operators representing local particle currents read
\begin{align}
j^\parallel_{l,r}&=iJ\left(b_{l,r}^\dagger b_{l,r+1}-\text{h.c.}\right)\label{def_local_curr_parr},\\
j^\perp_{r}&= iJ_\perp\left(e^{-ir\phi}b_{0,r}^\dagger b_{1,r}-\text{h.c.}\right)\label{def_local_curr_perp}.
\end{align}
Here, $j^\parallel_{l,r}$  represents the particle current from site $(l,r+1)$ to site $(l,r)$ and $j^\perp_{r}$ represents the particle current from site $(1,r)$ to site $(0,r)$.
Furthermore, the chiral-edge-current operator accounting for global particle transport along the legs of the system is given by
\begin{align}
j_c=\frac{1}{L}\sum_{r=0}^{L-1}\left(j^\parallel_{0,r}-j^\parallel_{1,r}\right).
\end{align}
\subsubsection{Integrated leg-gauge  momentum-distribution function}
\label{sec:mdfs0}
(Quasi-)momentum-distribution functions exhibit characteristics specific to certain zero-temperature phases of the ladder Hamiltonian, Eq.~(\ref{ham_2legladder_runggauge}) (see the discussion in \cite{paredes_2014,greschner_2016}).
Moreover, they can  in principle be obtained from experimental time-of-flight measurements \cite{Bloch2008}.
%

%
The specific momentum-distribution function (MDF) $n(k)$ that we primarily consider in this paper is defined by
\begin{align}
n\left(k_m\right) = \left\langle \tilde{b}_{0,k_m}^\dagger \tilde{b}_{0,k_m} + \tilde{b}_{1,k_m}^\dagger \tilde{b}_{1,k_m} \right\rangle,
\end{align}
using the leg-gauge operators of Eq.~(\ref{leg-gauge_mom_op}). We will refer to  $ n(k)$  as {\it integrated leg-gauge MDF}. 
The expectation value $\langle\bullet\rangle$ can either refer to the ground state or to a finite-temperature ensemble.
\subsubsection{Leg-resolved MDFs}
\label{sec:mdfs}
In the context of the ladder model, Eq.~(\ref{ham_2legladder_runggauge}), other distribution functions that are related to  $n(k)$ have been considered~\cite{paredes_2014,citro_2018,strinati_2018}, as follows.
(i) Leg-resolved leg-gauge MDFs $ n_l(k)$ are given in terms of leg-gauge momentum operators,
\begin{align}
n_l\left(k_m\right) = \left\langle \tilde{b}^\dagger_{l,k_m} \tilde{b}_{l,k_m} \right\rangle \,.
\end{align}
Summing over both legs, they add up to the integrated leg-gauge MDF, $ n(k) =  n_0(k)+ n_1(k)$.
%

%
(ii) Leg-resolved rung-gauge MDFs $\overline{n}_l(k)$ are defined by means of rung-gauge momentum operators $\overline{b}_{l,k_m}^{(\dagger)}$,
\begin{align}
\overline{n}_l\left(k_m\right) &= \left\langle  \overline{b}^\dagger_{l,k_m} \overline{b}_{l,k_m} \right\rangle \,,&
\overline{b}_{l,k_m} &= \frac{1}{\sqrt{L}}\sum_{r=0}^{L-1}e^{i k_m r}b_{l,r} \,.
\end{align}
We refer to  these operators $b_{l,r}^{(\dagger)}$, which are also used in Eq.~(\ref{ham_2legladder_runggauge}),  as rung-gauge operators, indicating that the  complex hopping amplitudes of the Hamiltonian are located along the rungs of the ladder.
The integrated rung-gauge MDF is given by $\overline{n}(k)=\overline{n}_1(k)+\overline{n}_2(k)$.
%

\begin{figure}
	\includegraphics[]{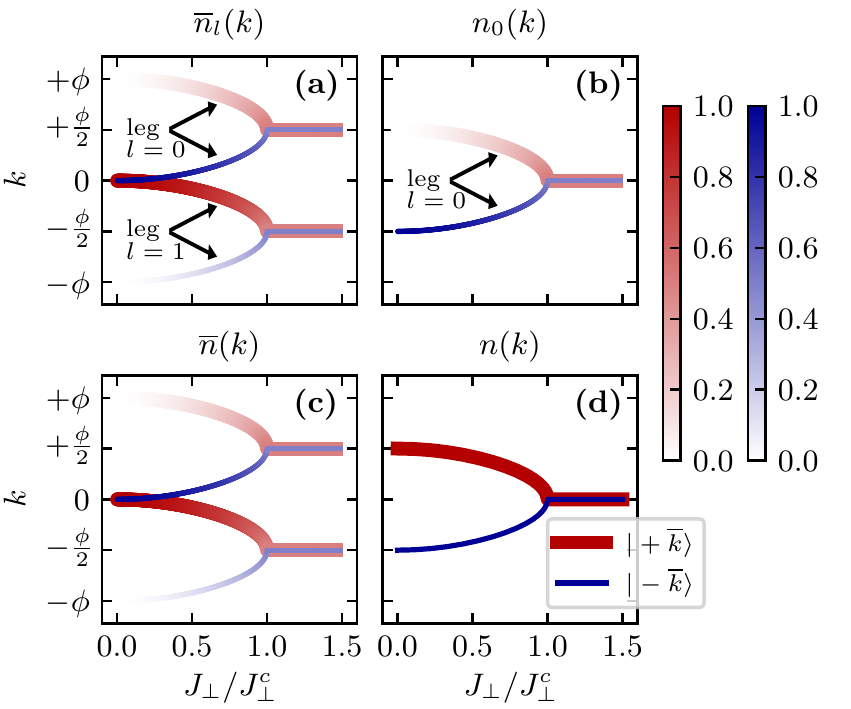}
	\caption{
		{\it Comparison of momentum-distribution functions in the single-particle case.}
		The figure shows the position of the maxima in various momentum-distribution functions that are (a),(b) leg-resolved ($l=0,1$) or (c),(d) integrated over $l$.
		The values of momentum-distribution functions for each of the single-particle ground states $\ket{\pm\overline{k}}$ are indicated by the colorbars.
 		Panels (a) and (c) depict the behavior in the rung gauge, while (b) and (d) are for the leg gauge.
 		The integrated leg-gauge momentum-distribution function $n(k)$ of panel (d) 
		will be discussed in the remainder of the paper. See the text in Secs.~\ref{sec:mdfs0} and \ref{sec:mdfs}.	%
	}
	\label{diff_gauges}
\end{figure}

%
Leg-resolved leg-gauge and leg-resolved rung-gauge MDFs are related as follows:
\begin{align}
n_0(k)&=\overline{n}_0\left(k+\frac{\phi}{2}\right),&
n_1(k)&=\overline{n}_1\left(k-\frac{\phi}{2}\right).\label{rel_rung_leg_mom}
\end{align}
We compare all four types of MDFs (integrated versus leg resolved; leg gauge versus rung gauge) in Figs.~\ref{diff_gauges}(a)-\ref{diff_gauges}(d).
Generally, the number of possible maxima is doubled by going from leg to rung gauge.
In the experiment in Ref.~\cite{atala_2014} the rung gauge was realized which leads to the MDF $\overline{n}(k)$.
For convenience, we will present results for the leg gauge, i.e., $n(k)$, yet the former can be obtained from the latter.
The analysis of signatures of the vortex-fluid phase proposed in Sec.~\ref{sec:contrast} can straightforwardly be adapted to $\overline{n}(k)$ by simply restricting $k$ to one-half of the Brillouin zone. 
%
%
%
%
\section{Canonical purification}
\label{sec_purification}
We here provide details of our finite-temperature matrix-product state methods that will be used for hard-core bosons, putting a particular focus on canonical simulations.
The reader primarily interested in the physics may skip this part and jump directly to Sec.~\ref{main_results}.
%

%
The core concept of purification is that a mixed state in a physical space can be represented as the partial trace of a pure state in an artificially extended space~\cite{schollwock2011density}.
Thus, by incorporating an auxiliary counterpart for each physical site, thermal states can be represented as matrix-product states and propagated in imaginary time using well-developed techniques~\cite{daley_kollath_sw_vidal_2004,bruognolo_2017,hauschild2017finding,barthel2016matrix,pollmann_2015,tiegel_2014,alvarez_2013,karrasch_2013,karrasch_2012,stoudenmire_white_2010,barthel_sw_white_2009,mcculloch2007density,feiguin_2005,Verstraete_2004,Bursill_1996}.
In this context, suitable initial infinite-temperature states need to be chosen.
For grand-canonical calculations these states are straightforward to generate~\cite{schollwock2011density}: they exhibit maximum entanglement between each physical site and its auxiliary counterpart but are otherwise of product form. 
On the other hand, the required initial states for canonical simulations exhibit nontrivial long-range correlations.
They have been successfully obtained as the ground states of specifically chosen Hamiltonians using the density-matrix renormalization-group method~\cite{feiguin_2010,nocera_2016,Koehler2018}.
%

%
Here, we generate matrix-product-state representations of canonical infinite-temperature states by employing a bookkeeping mechanism that explicitly accounts for the occupation of physical sites within the matrix-product state. 
This approach renders the generation of canonical infinite-temperature states straightforward.
We emphasize that the methodology employed here is very similar to the one introduced in Ref.~\cite{barthel2016matrix}.
%

%
Our purification approach can be subdivided into three steps.
(i) First, a matrix-product-state representation of the (grand-)canonical infinite-temperature state is constructed.
%
(ii) Second, a suitable matrix-product-operator representation of the propagator $\exp\left(-\beta H \right)$ is constructed, where $\beta=1/T$ is the inverse temperature.
%
(iii)  A third step involves the imaginary-time propagation, including the control of errors, to obtain a thermal state corresponding to a finite temperature.
%

%
In the remainder of this section, we will describe all steps in greater detail.
\subsection{Matrix-product-state structure}
\begin{figure}
	\includegraphics[]{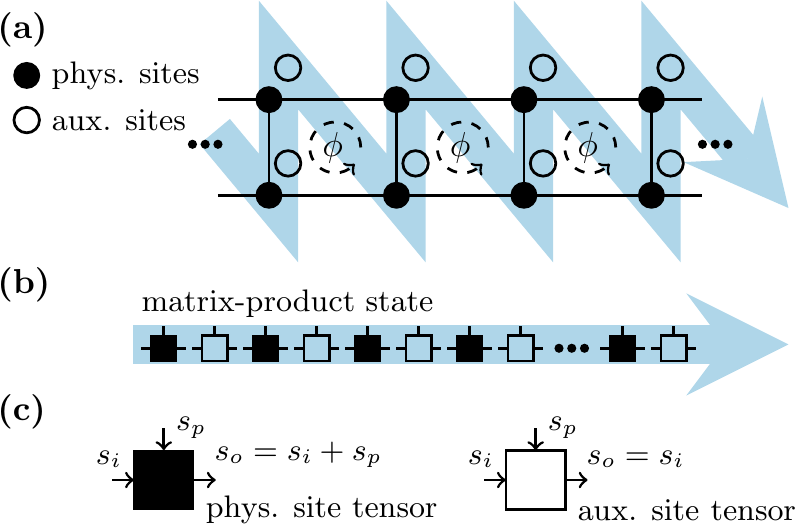}
	\caption{
		{\it Purification approach using matrix-product states.}
		Visualization of the matrix-product-state structure employed during (grand-)canonical purification. 
		(a) Sketch of the ladder Hamiltonian.
		The bonds (solid lines) represent the Hamiltonian terms and filled circles represent the physical lattice sites. 
		Within the purification approach, each physical site is accompanied by an auxiliary counterpart represented by an empty circle.
		(b) The matrix-product state is comprised of tensors representing physical sites (filled squares) and of tensors representing auxiliary sites (empty squares).
		These tensors are aligned side by side following a zigzag pattern through the ladder as indicated by the blue arrows in (a) and (b).
		(c) As discussed in the main text, the canonical purification approach requires the bookkeeping of the total occupation of all physical sites.
		Therefore, the incoming left leg $i$ of each tensor is split into blocks $\left\lbrace s_i \right\rbrace$ referring to different occupations of the physical sites to the left of the specific tensor.
		Similarly, the incoming leg $p$ entering each tensor from above is split into blocks $\left\lbrace s_p \right\rbrace$ referring to the occupation of the specific site represented by the tensor, and the outgoing leg $o$ to the right of each tensor is split into blocks $\left\lbrace s_o \right\rbrace$ referring to the occupation of all physical sites to the left of the specific tensor including the tensor itself.
	}
	\label{puri_mps}
\end{figure}

The matrix-product states employed during (grand-)canonical purification simulations are comprised of tensors representing physical sites of the two-leg ladder system and their auxiliary counterparts; see Fig.~\ref{puri_mps}(a).
As shown in Fig.~\ref{puri_mps}(b), these tensors are aligned side by side following a zigzag pattern through the ladder and forming a chain of length $4L$.
%

%
Since canonical calculations are at a fixed number of particles, which is here conserved by the ladder Hamiltonian, we need the matrix-product state to account for the corresponding symmetry.
In practice, this means that, on the level of each tensor, a bookkeeping mechanism needs to be implemented.
This mechanism keeps track of the total occupation number of the physical sites when going through the matrix-product state from left to right.
At this point, we recap that the conventional building blocks of a matrix-product state are three-legged tensors, here denoted by $\mathcal{T}^{p,i}_o$. 
The incoming leg $p$, which is usually referred to as \emph{physical leg}, corresponds to local basis states of the lattice site represented by the tensor.
The incoming leg $i$ and the outgoing leg $o$ enable a connection between the tensor and its neighbors within the matrix-product state.
Against this background, the employed bookkeeping mechanism can be understood as a subdivision of the vector spaces associated with each of the conventional legs into sectors corresponding to different occupation numbers.
More precisely, for each tensor, the sectors of its incoming leg $i$ correspond to occupation numbers of the physical sites that are represented by the tensors to its left.
Analogously, for each tensor, the sectors of its outgoing leg $o$ correspond to occupation numbers of the physical sites that are represented by the tensor itself and by the tensors to its left.
Moreover, the sectors subdividing an incoming leg $p$ refer to the local occupation numbers of the basis states.
Hence the outgoing leg $o$ of the leftmost tensor in the matrix-product state contains information about the total occupation of all physical sites.
This bookkeeping mechanism is illustrated in Fig.~\ref{puri_mps}(c).
\subsection{Canonical infinite-temperature state}
Matrix-product operators are required for the preparation of the (grand-)canonical infinite-temperature state as well as for the subsequent imaginary-time propagation.
Here, they are constructed from single-site operators following the generic approach presented in Ref.~\cite{hubig_2017}.
Considering the  ladder model Eq.~(\ref{ham_2legladder_runggauge}) in the hard-core boson limit, a possible grand-canonical infinite-temperature state $\ket{\beta=0}_\text{g.c.}$ can be readily constructed using single-site operators ${b_{l,r}^{(\dagger)}}_\text{phys.}$ and ${b_{l,r}^{(\dagger)}}_\text{aux.}$ which act on physical sites and their auxiliary counterparts.
This state is spread over all possible sectors corresponding to total occupation numbers of the physical sites ranging from zero to $2L$. 
It is given by
\begin{align}
\ket{\beta=0}_\text{g.c.} = 
\bigotimes \limits_{l,r} \frac{1}{\sqrt{2}}
\left(
{b_{l,r}^{\dagger}}_{\text{phys.}} + {b_{l,r}^{\dagger}}_{\text{aux.}}
\right)
\ket{\text{vac}},
\end{align}
with $\ket{\text{vac}}$ being the vacuum state.
We emphasize that a suitable canonical infinite-temperature state with a fixed total occupation number can be obtained from $\ket{\beta=0}_\text{g.c.}$ by means of a projection onto the corresponding particle number sector and subsequent normalization.
Moreover, the matrix-product-state structure introduced above renders this projection straightforward: all but the symmetry sector of interest are set to zero at the very right end of the matrix-product state.
Finally, the canonical infinite-temperature state with a fixed total occupation number of $N$ particles is given by
\begin{align}
\ket{\beta=0}_{\text{c}} = \mathcal{P}_N \ket{\beta=0}_\text{g.c.},
\end{align}
where $\mathcal{P}_N$ denotes a projection onto the corresponding subspace and subsequent normalization of the matrix-product state.
\subsection{Imaginary-time propagation}
The imaginary-time evolution requires a matrix-product-operator representation of the propagator $e^{-\beta H}=\left(e^{-\tau H}\right)^{\beta/\tau}$.
For this, we subdivide the ladder Hamiltonian given in Eq.~(\ref{ham_2legladder_runggauge}) into two parts, $H=H_0+H_1$, and approximate the time-evolution operator by means of a second-order Trotter decomposition,
\begin{align}
e^{-\tau H} = e^{-\tau H_0/2}e^{-\tau H_1}e^{-\tau H_0/2}+\mathcal{O}\left(\tau^3\right).\label{trotter_decomp}
\end{align}
One possible choice for  $H_0$ and $H_1$ is given by ($j=1,2$)
\begin{align}
H_j=&
-\sum\limits_{r=0}^{L-2} \mathcal{I}_j^r
\left(
J\sum_{l=0}^{1}b_{l,r}^\dagger b_{l,r+1}
+J_\perp e^{-ir\phi}b_{0,r}^\dagger b_{1,r}
\right)\nonumber\\
&- \mathcal{I}_j^L J_\perp e^{-i(L-1)\phi}b_{0,L-1}^\dagger b_{1,L-1} + \text{h.c.},\label{ham_division}
\end{align}
with $\mathcal{I}_m^n = (m+n)\mathop{\text{mod}}2$.
Note that for hard-core bosons ($U/J\rightarrow\infty$), the interaction term in Eq.~(\ref{ham_2legladder_runggauge}) is implemented by means of the cutoff of the lattice site-local Hilbert space at a maximum of one boson.
The Hamiltonian parts $H_0$ and $H_1$ are composed of commuting contributions which can be associated with every second plaquette of the ladder, meaning that they can be diagonalized numerically. 
Hence the matrix-product operators representing the exponentials in Eq.~(\ref{trotter_decomp}) can be constructed automatically from single-site operators using common matrix-product-operator arithmetic~\cite{hubig_2017}.
In terms of the employed matrix-product-state structure, the Hamiltonian $H$ exhibits at most fourth-nearest-neighbor interactions and the propagators $e^{-\tau H_0}$ and $e^{-\tau H_1}$ exhibit at most sixth-nearest-neighbor interactions.
%

%
The imaginary-time propagation itself consists of the sequential application of matrix-product operators representing the propagators on the right-hand side of Eq.~(\ref{trotter_decomp}).
Choosing a suitable step width $\tau$, they are applied repeatedly to the initial (grand-)canonical infinite-temperature state $\ket{\beta=0}_\text{c}$ ($\ket{\beta=0}_\text{g.c.}$), which decreases the temperature of the so-evolved thermal state.
Between these applications, the bond dimensions of the evolved matrix-product state need to be truncated and, due to the fact that the imaginary-time propagators are not unitary, the matrix-product state also needs to be repeatedly normalized.
During the imaginary-time propagation, errors arise due to the Trotter decomposition of the propagator and due to the repeated truncation of the evolved state.
Regarding the numerical results presented in this paper, these errors have been controlled independently by comparing results obtained for different truncation thresholds at fixed Trotter-step widths and vice versa.
We note that this error control can in principle be based on the monitoring of state overlaps.
However, here it is sufficient to focus on the values obtained for the actual observables of interest, which are chiral edge currents and momentum-distribution functions.
%
%
%
%
\section{Vortex-fluid-to-Meissner crossover at finite temperatures}
\label{main_results}
In Sec.~\ref{boson_results}, we address the grand-canonical statistics of the noninteracting bosonic ladder model.
Focusing on momentum-distribution functions, the presence of clearly detectable characteristics of the vortex-fluid phase at finite temperatures is examined.
In Sec.~\ref{ferm_results}, we concentrate on the grand-canonical statistics of noninteracting spinless fermions and elucidate the differences between the bosonic and fermionic vortex(-fluid)-to-Meissner transition.
Our main results regarding the vortex-fluid-to-Meissner crossover of canonical and grand-canonical thermal states of hard-core bosons are presented in Sec.~\ref{hcb_results}.
\subsection{Noninteracting bosons}
\label{boson_results}
We recap that the noninteracting bosonic ladder model with periodic boundary conditions exhibits two degenerate ground states in the vortex-fluid phase, corresponding to finite characteristic momenta $\pm\overline{k}$ (see Sec.~\ref{sec:nonint}).
At the vortex-fluid-to-Meissner transition, this degeneracy is lifted and the system exhibits a unique ground state corresponding to a vanishing characteristic momentum, $k=0$, in the Meissner phase.
\subsubsection{Chiral edge currents}
For ladder systems with periodic boundary conditions, the chiral edge current $\bra{\overline{k}_0}j_c\ket{\overline{k}_0}$ corresponding to the unique bosonic ground state in the Meissner phase, denoted by $\ket{\overline{k}_0}$, does not vary with increasing interleg coupling strength, $\partial_{J_\perp}\bra{\overline{k}_0}j_c\ket{\overline{k}_0}=0$, and shows zero fluctuations, $\bra{\overline{k}_0}j_c^2\ket{\overline{k}_0}-\left(\bra{\overline{k}_0}j_c\ket{\overline{k}_0}\right)^2=0$.
This does not apply to the degenerate ground states found in the vortex-fluid phase.
Moreover, in the vortex-fluid phase, each of the degenerate ground states exhibits a population imbalance between the legs of the ladder.
%

%
\begin{figure}
	\includegraphics[]{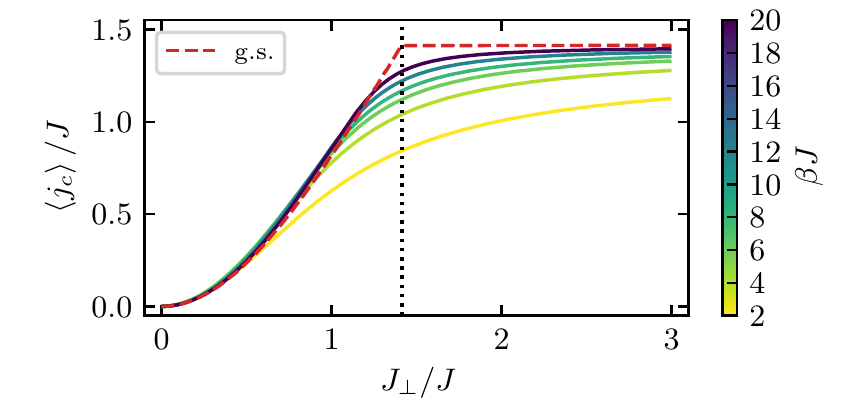}
	\caption{
		{\it Noninteracting bosons, $f=1/2$, $\langle j_c\rangle$.} Chiral edge currents $\left\langle j_c \right\rangle$ corresponding to noninteracting bosons as a function of interleg coupling strength $J_\perp$.
		Considering a ladder of length $L=128$ with open boundary conditions and grand-canonical thermal states corresponding to different inverse temperatures $\beta=2,4,6,8,12,20/J$ (also indicated by the colorbar) at average filling one-half.
		The chiral edge current corresponding to the ground state (g.s.) at filling $f=1/2$ is shown for comparison.
		The dotted line indicates the zero-temperature critical $J_\perp/J$.
	}
	\label{chiral_currs_nonint_bosons_L_128}
\end{figure}
The red dashed line in Fig.~\ref{chiral_currs_nonint_bosons_L_128} shows the chiral edge current corresponding to the ground state of noninteracting bosons confined on a ladder of length $L=128$ with open boundaries at filling one-half.
At this point, we make clear that the filling, which is here also referred to by $f$, denotes the number of particles divided by $2L$, i.e., the number of  lattice sites.
The chiral edge current increases gradually with increasing interleg coupling strength in the vortex-fluid phase and remains constant in the Meissner phase.
A kink  in $\langle j_c\rangle$ reveals the critical interleg coupling strength of the vortex-fluid-to-Meissner transition.
We note that boundary and finite-size effects are negligible on the scale of the figure (by comparison to results for other $L$, not shown here).
Solid lines show the chiral edge currents of grand-canonical thermal states corresponding to different temperatures (colorbar) and average filling one-half.
The kink is quickly washed out and thus there is no clear signature of the crossover in the finite-temperature chiral edge currents.
Note that the chiral edge current can be slightly enhanced due to the effect of finite temperatures in the vortex-fluid phase.
\subsubsection{Integrated leg-gauge momentum-distribution functions}
\begin{figure}
	\includegraphics[]{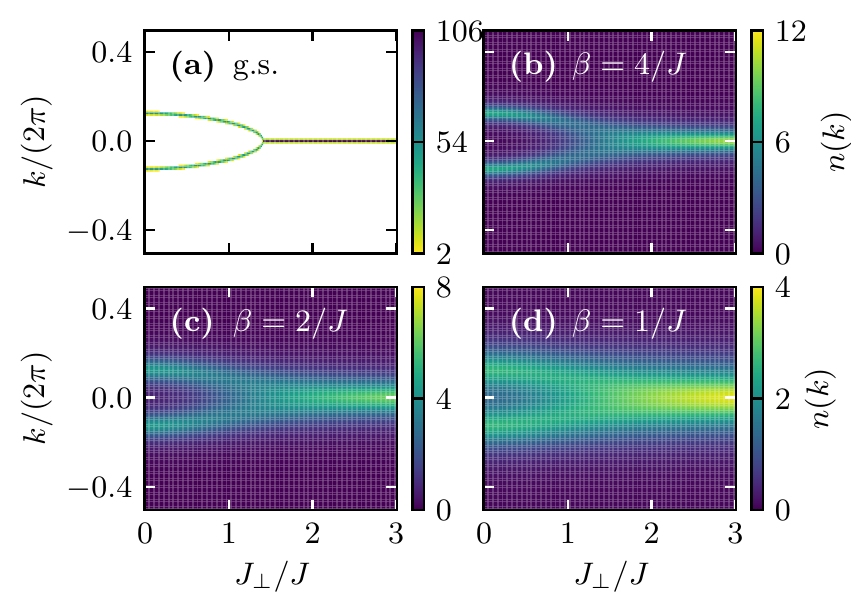}
	\caption{
		{\it Noninteracting bosons, $f=1/2$, $n(k)$.} Integrated leg-gauge momentum-distribution function $ n(k)$ as a function of interleg coupling strength $J_\perp$ for $L=128$ and open boundary conditions.  
		(a) Ground-state (g.s.) momentum-distribution functions at filling $f=1/2$.
		(b) - (d)  Momentum-distribution functions computed in  grand-canonical thermal states at average filling one-half and different inverse temperatures (b) $\beta=4/J$, (c) $\beta=2/J$, and (d) $\beta=1/J$.
	}
	\label{mom_dists_nonint_bosons_L_128}
\end{figure}

Results for the integrated leg-gauge MDF of noninteracting bosons are shown in Fig.~\ref{mom_dists_nonint_bosons_L_128}.
In this context, the ground-state vortex-fluid-to-Meissner transition manifests itself in the transition from two distinct maxima at finite momenta to a single maximum at zero momentum, as shown in Fig.~\ref{mom_dists_nonint_bosons_L_128}(a).
The integrated leg-gauge MDFs calculated in  grand-canonical thermal states at different temperatures $J/4$, $J/2$ and $J$, which are shown in Figs.~\ref{mom_dists_nonint_bosons_L_128}(b), \ref{mom_dists_nonint_bosons_L_128}(c), and \ref{mom_dists_nonint_bosons_L_128}(d), respectively,  reveal that the typical peaks persist at finite temperatures.
We note that the results for ground states in Fig.~\ref{mom_dists_nonint_bosons_L_128}(a) and for the grand-canonical thermal states in Figs.~\ref{mom_dists_nonint_bosons_L_128}(b), \ref{mom_dists_nonint_bosons_L_128}(c), and \ref{mom_dists_nonint_bosons_L_128}(d) are for a fixed filling $f=1/2$ and an average filling of one-half,  respectively.
\subsubsection{Signatures of the vortex-fluid phase at finite temperatures}
\label{sec:contrast}
\begin{figure}
	\includegraphics[]{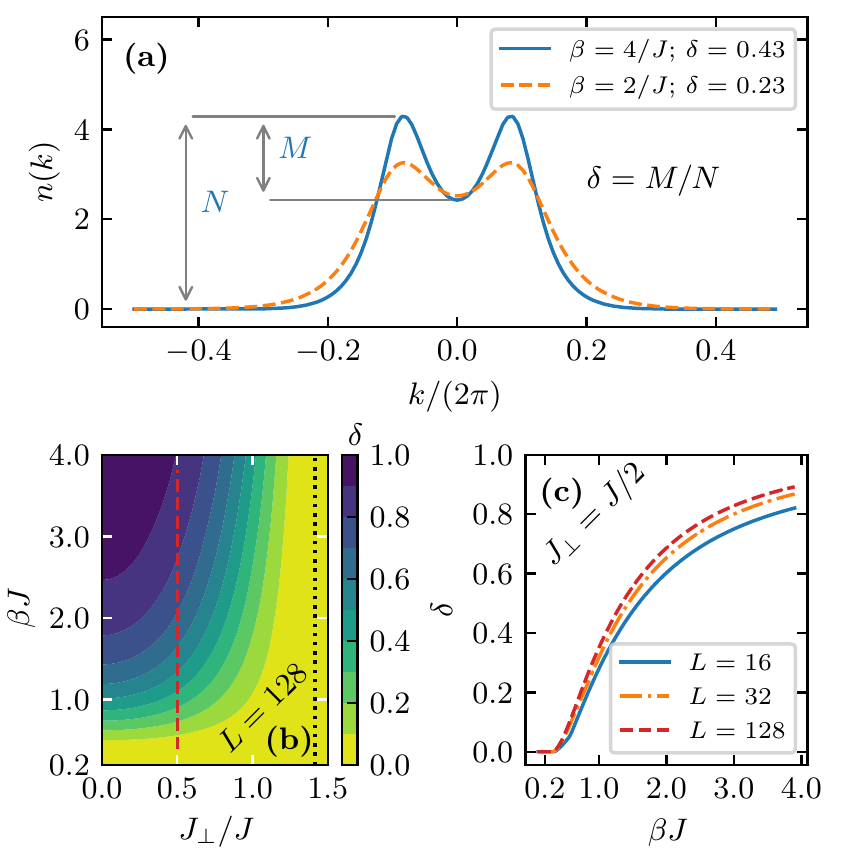}
	\caption{
		{\it Signatures of the vortex-fluid  phase of noninteracting bosons at $T>0$.}
		(a) Grand-canonical integrated leg-gauge momentum-distribution functions $n(k)$ at average filling one-half corresponding to inverse temperatures $\beta = 4/J$ and $\beta = 2/J$ at $J_\perp=J$.
		The two characteristic peaks indicating the vortex-fluid phase are more distinctive at lower temperatures. 
		The contrast $\delta$ as defined in Eq.~(\ref{delta}) serves as an indicator for the vortex-fluid phase.
		(b) Contrast $\delta$ as a function of interleg coupling strength $J_\perp$ and inverse temperature $\beta$ for grand-canonical thermal states at average filling one-half.
		The dotted line indicates the position of the zero-temperature transition occurring at $J_\perp=\sqrt{2}J$.
		(c) Illustration of finite-size effects:  $\delta$ versus $\beta J$ for ladders of length $L=16$, $32$ and $128$. 
		Panel (c) is a cut  through  panel (b) at $J_\perp=J/2$ [see the dashed line in (b)].
	}
	\label{delta_nonint_bosons_L_128}
\end{figure}

Next, we elucidate the finite-temperature vortex-fluid-to-Meissner crossover.
Considering the integrated leg-gauge MDF $n(k)$ of thermal states, we specify at which temperatures clear characteristics of the vortex-fluid phase can be detected and where the crossover can be observed.
For this purpose, we introduce a measure for the contrast of finite-momentum maxima in $n(k)$ via this definition 
\begin{align}
\delta = \frac{\max_k  [n(k)]  - n(k=0)}{\max_k [ n(k)] }\,.\label{delta}
\end{align}
Thus, on the one hand, values of $\delta$ larger than zero indicate that two characteristic peaks at finite momenta can be resolved. 
Hence $\delta>0$ is indicative of  the vortex-fluid phase.
On the other hand, values of $\delta$ equal to zero mean that $n(k)$ exhibits a single maximum at zero momentum, and, due to the blurring effect of finite temperatures, it cannot be decided whether the underlying values of $J$, $J_\perp$ and $\phi$ correspond to the vortex-fluid or Meissner phase.
The definition of the contrast, Eq.~(\ref{delta}), is also illustrated in Fig.~\ref{delta_nonint_bosons_L_128}(a) using the abbreviations $M=\max_k  [n(k)]  - n(k=0)$ and $N=\max_k [ n(k)]$.
There, it is exemplified that the characteristic peaks indicating the vortex-fluid phase are more distinctive at lower temperatures.
Considering grand-canonical thermal states at average filling one-half, a contour plot of $\delta$ as a function of interleg coupling $J_\perp$ and  inverse temperature $\beta$ is shown in  Fig.~\ref{delta_nonint_bosons_L_128}(b).
It becomes apparent that the integrated  leg-gauge MDFs show clear signatures of the vortex-fluid phase at sufficiently small values of $J_\perp \lesssim J $ and  up to quite high  temperatures $T\sim J$. 
Note that the zero-temperature transition occurs at $J_\perp=\sqrt{2}J$, as indicated in Fig.~\ref{delta_nonint_bosons_L_128}(b) by the dotted line.
The role of finite-size effects for the contrast $\delta$ is studied in Fig.~\ref{delta_nonint_bosons_L_128}(c) by considering a cut through the graph in Fig.~\ref{delta_nonint_bosons_L_128}(b) at $J_\perp=J/2$.
We note that finite-size effects of $\delta$ are overall small and, for a ladder of length $L=128$, the differences to larger $L$ (not shown in the figure) are negligible on the scale of the plot. 
Furthermore, we only observe quantitative differences as $L$ varies, yet the overall dependence of $\delta$ on $\beta$ is the same for all $L$.
%

%
Finally, we comment on another measure for the vortex phases that is often studied in the literature (see, e.g., Refs.~\cite{giamarchi_2001,piraud_2015,greschner_2016}), namely the vortex density $l_v$.
In Ref.~\cite{piraud_2015}, $l_v$ was extracted from a Fourier transformation of the local rung-current patterns. 
At finite temperatures, however, the local rung currents are quickly washed out such that this approach cannot be used here. 
Rather, one would need to study the decay of rung-current equal-time autocorrelations. 
These  are not easily accessible in experiments and we therefore do not further study measures for the vortex density here.

\subsection{Noninteracting spinless fermions}
\label{ferm_results}
In this section, we discuss chiral edge currents and momentum-distribution functions of noninteracting spinless fermions.
Note that the chiral current as a function of flux per particle was studied in Ref.~\cite{strinati_2017}.
\subsubsection{Chiral edge currents}
\label{sec:jc_fermions}
\begin{figure}
	\includegraphics[]{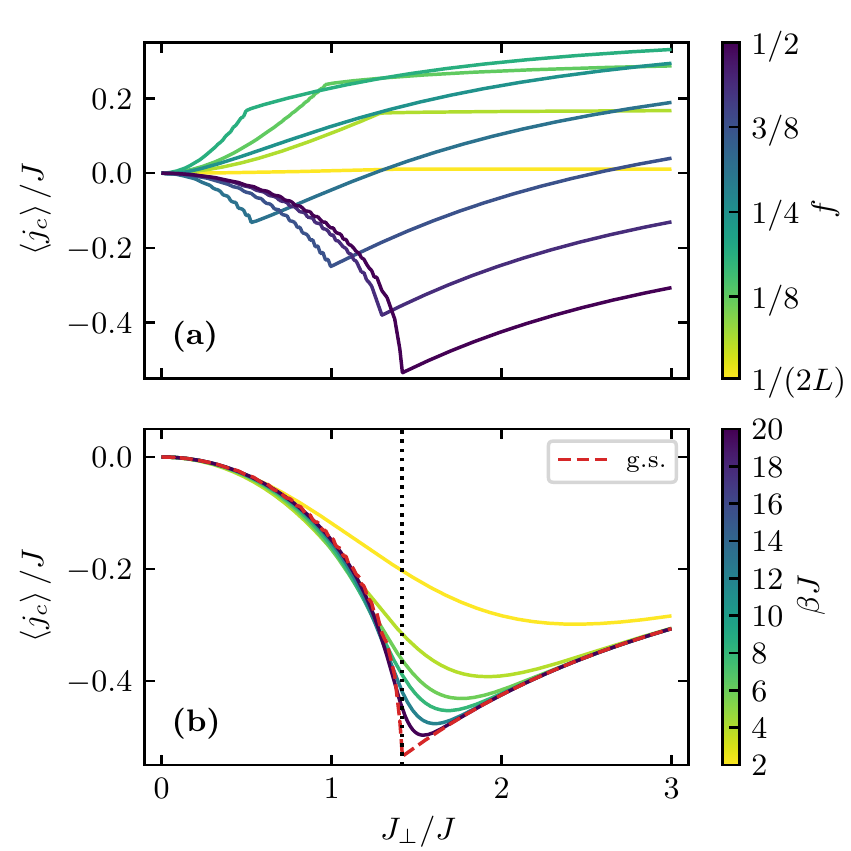}
	\caption{
		{\it Noninteracting fermions, $\langle j_c \rangle$.} Chiral edge currents $\left\langle j_c \right\rangle$  corresponding to noninteracting spinless fermions as a function of interleg coupling strength $J_\perp$ for  $L=128$ and open boundary conditions.
		(a) Zero temperature $T=0$: the ground-state results for different fillings $1/(2L)\le f\le 1/2$ (values of $f$ are indicated by the colorbar) reveal a current reversal with increasing $f$.
		At filling $f=1/2$, the cusp of the ground-state chiral edge current clearly indicates the vortex-to-Meissner transition at $J_\perp = \sqrt{2}J$.
		(b) Results for $\langle j_c \rangle$ computed in grand-canonical thermal states for different inverse temperatures $\beta$ (colorbar) at average filling one-half.
		The chiral edge current corresponding to the ground state (g.s.) at filling one-half is shown for comparison (dashed line).
		The dotted line indicates the zero-temperature critical $J_\perp/J$ at $f=1/2$.
	}
	\label{chiral_currs_nonint_fermions_L_128}
\end{figure}

Chiral edge currents corresponding to the ground states of noninteracting spinless fermions behave fundamentally different from those corresponding to the ground states of noninteracting bosons.
For bosons, the behavior close to the band minima is important, while for fermions the behavior close to the Fermi energy is relevant, or for nonuniversal objects, such as the chiral edge current, the occupation of all quasimomenta~\cite{carr_2006}.
One notable consequence is that already for $U=0$, in the Meissner phase, noninteracting fermions feature a reversal of the direction of the chiral edge current with increasing filling $f$~\cite{carr_2006,roux_2007}.
Fermionic ground-state chiral edge currents are shown in Fig.~\ref{chiral_currs_nonint_fermions_L_128}(a) as a function of interleg coupling $J_\perp$ for different fillings $f$ ranging from $1/(2L)$ to $1/2$.
For each filling, a kink of the chiral edge current clearly reveals the critical value of $J_\perp$ at the vortex-to-Meissner transition.
Figure~\ref{chiral_currs_nonint_fermions_L_128}(b) concentrates on fermionic chiral edge currents at finite temperatures and average filling one-half.
For temperatures below approximately $J/10$ and values of $J_\perp \lesssim J$ or $ J \gtrsim 2J$, i.e., deep within the Meissner and vortex phases, the chiral edge currents computed in  grand-canonical thermal states coincide with the ones obtained in the fermionic ground state. 
However, even for the smallest temperature considered in Fig.~\ref{chiral_currs_nonint_fermions_L_128}(b), which is $J/20$, the chiral edge current does not exhibit a clear kink at the critical value, $J_\perp=\sqrt{2}J$, of the vortex-to-Meissner transition.
We note that the minima of the chiral edge current systematically overestimate the critical value of $J_\perp$ with increasing temperature.
\subsubsection{Integrated leg-gauge momentum-distribution functions}
\begin{figure}
	\includegraphics[]{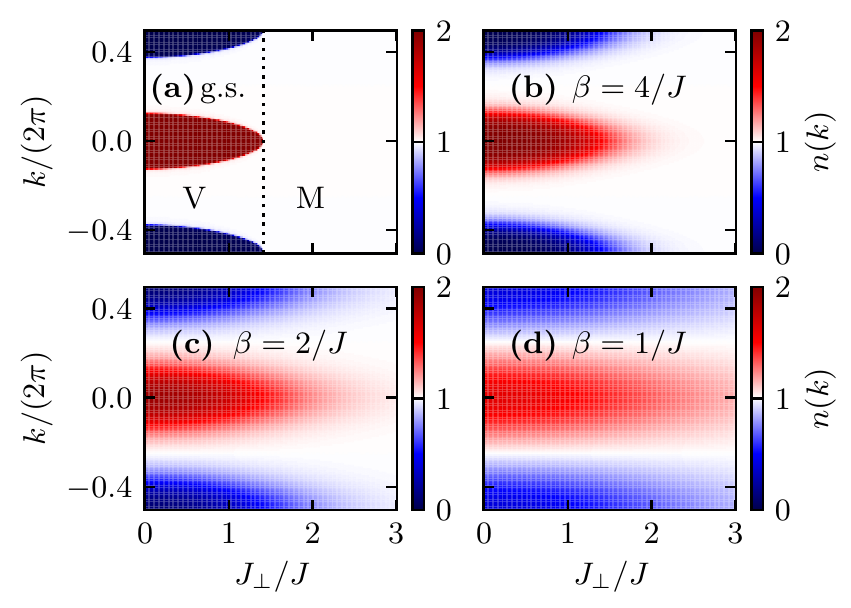}
	\caption{
		{\it Noninteracting fermions, $f=1/2$, $n(k)$.} Integrated leg-gauge momentum-distribution functions $n(k)$ as a function of interleg coupling strength $J_\perp$ for a ladder of length $L=128$ with open boundary conditions.
		(a)  Ground-state (g.s.) momentum-distribution function at filling $f=1/2$.
		In the vortex phase~(V), the ground-state momentum-distribution function shows a maximum value of two at certain momenta due to the occupation of the upper $\epsilon_{-}$ band.
		On the other hand, in the Meissner phase~(M), the lower $\epsilon_{+}$ band is energetically well separated from the upper $\epsilon_{-}$ band. 
		Thus, at filling $f=1/2$, it is fully occupied.
		This leads to a constant momentum-distribution function $ n(k) = 1$.
		The dotted line indicates the zero-temperature critical $J_\perp/J$.
		(b)--(d) Momentum-distribution functions computed in grand-canonical thermal states at average filling one-half and for different inverse temperatures (b) $\beta=4/J$, (c) $\beta=2/J$, and (d) $\beta=1/J$.
	}
	\label{mom_dists_nonint_fermions_L_128}
\end{figure}

Within the Meissner phase, the integrated leg-gauge MDFs corresponding to the ground state of noninteracting spinless fermions at filling $f=1/2$ take a constant value equal to one; see Fig.~\ref{mom_dists_nonint_fermions_L_128}(a) (note that white corresponds to an occupation of one in the color-coding of that figure).
This can be readily understood from the overlap of the bands referred to in Eq.~(\ref{band_overlap}) and shown in Fig.~\ref{nonint}(c).
In the Meissner phase, the lower band is energetically well separated from the upper band.
Hence, at filling one-half, it is fully occupied which leads to a constant MDF.
Note that the bands shown in Fig.~\ref{nonint} and referred to in Eq.~(\ref{band_overlap}) correspond to the ladder model with periodic boundary conditions while we show data
for open boundary conditions in the figure.
However, for systems with a large number of rungs $L$, boundary effects play a minor role and integrated leg-gauge MDFs for open and periodic boundary conditions  coincide.
On the other hand, in the vortex phase, the ground-state integrated leg-gauge MDF is piecewise constant and takes on the values zero, one, and two, corresponding to nonoccupied, singly occupied and doubly occupied momenta; see Fig.~\ref{mom_dists_nonint_fermions_L_128}(a).
With the onset of finite temperatures, the sharp features of the fermionic ground-state integrated leg-gauge MDF blur out and their discreteness is lost.
Figures~\ref{mom_dists_nonint_fermions_L_128}(b), \ref{mom_dists_nonint_fermions_L_128}(c) and \ref{mom_dists_nonint_fermions_L_128}(d) show integrated leg-gauge MDFs computed in grand-canonical thermal states at average filling one-half and at different temperatures $J/4$, $J/2$, and $J$, respectively.
In the case of fermions, finite temperature causes the peak structure characteristic for the vortex phase to extend into the Meissner region, opposite from the behavior of bosons.
\subsection{Hard-core bosons}
\label{hcb_results}
In this section, we give an account of our main results addressing the properties of hard-core bosons on the ladder system introduced in Eq.~(\ref{ham_2legladder_runggauge}) at finite temperatures.
We remind the reader that a value of $\phi=\pi/2$ is considered.
\subsubsection{Chiral edge currents}
Let us start with the presentation of our results concerning the chiral edge current $\left\langle j_c \right\rangle$.
In the thermodynamic limit, a kink at the maximum value of $\left\langle j_c \right\rangle$ indicates the critical value of the interleg coupling strength $J_\perp$ of the ground-state vortex-fluid-to-Meissner transition.
Furthermore, in the Meissner phase, $\left\langle j_c \right\rangle$ is expected to decay to zero as $\left\langle j_c\right\rangle\propto1/J_\perp$~\cite{piraud_2015}.
%

\begin{figure}
	\includegraphics[]{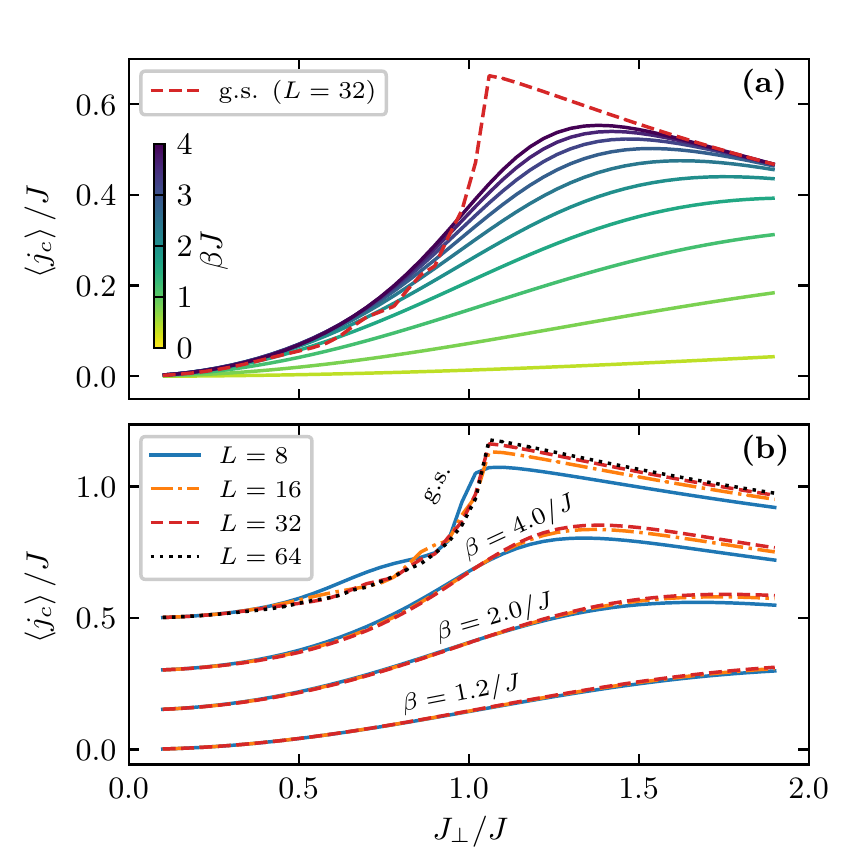}
	\caption{
		{\it Hard-core bosons, $f=1/2$, $\langle j_c \rangle$.} (a) Chiral edge currents $\left\langle j_c \right\rangle$  corresponding to canonical thermal states of hard-core bosons at filling $f=1/2$ as a function of interleg coupling strength $J_\perp$.
		We show results for a ladder of length $L=32$ and different inverse temperatures $\beta$ indicated by the colorbar.
		The corresponding ground-state chiral edge current (g.s.) is shown for comparison.
		(b) Illustration of  finite-size effects, comparing the chiral edge currents for ladders of length $L=8,16,32$ (for all $T\geq 0$) and $L=64$ (ground states only). 
		For better visibility, chiral-edge-current data for $\beta=2/J$, $\beta=4/J$ and the ground states (g.s.) are vertically offset by $0.15$, $0.3$ and $0.5$, respectively.	
}
	\label{chiral_currents_comb}
\end{figure}

%
Figure~\ref{chiral_currents_comb}(a) shows chiral edge currents corresponding to the ground states of a ladder system of length $L=32$ (red dashed line) at filling one-half.
For this value of $L$, the kink of the chiral edge current clearly reveals the location of the ground-state vortex-fluid-to-Meissner transition at $J_\perp\approx J$.
The solid colored lines in Fig.~\ref{chiral_currents_comb}(a) show chiral edge currents corresponding to canonical thermal states of different temperatures ranging from  $J/4$ to $5J/2$ at filling one-half; note that inverse temperatures $\beta$ are indicated by the colorbar.
For values of $J_\perp$ approximately greater than $1.5J$, i.e., deep within the Meissner phase, and for temperatures below around $J/3$, the chiral edge currents corresponding to the canonical thermal states overlap with the ones corresponding to the ground state. 
Further, for sufficiently small temperatures, the finite-temperature currents show clear maxima.
However, the positions of these maxima tend to systematically overestimate the critical value of $J_\perp$ with increasing temperatures.
%

%
Figure~\ref{chiral_currents_comb}(b) elucidates the role of finite-size effects. 
The  chiral edge currents for canonical thermal states with inverse temperatures $\beta=1.2/J$, $\beta=2/J$, $\beta=4/J$ at filling $f=1/2$ are shown for ladders of length $L=8$, $L=16$, $L=32$ and $L=64$ (the latter for the ground state only).
At zero temperature and in the Meissner phase, the chiral edge currents for the  $L=64$ system are slightly more pronounced than those for the $L=32$ system.
However, as can be seen in Fig.~\ref{chiral_currents_comb}(b), finite-size effects of the chiral edge currents play a  minor quantitative role, in particular at $T>0$, where there are small differences between the  $L=32$ and $L=16$ data for the selected values of $T$.
%

\begin{figure}
	\includegraphics[]{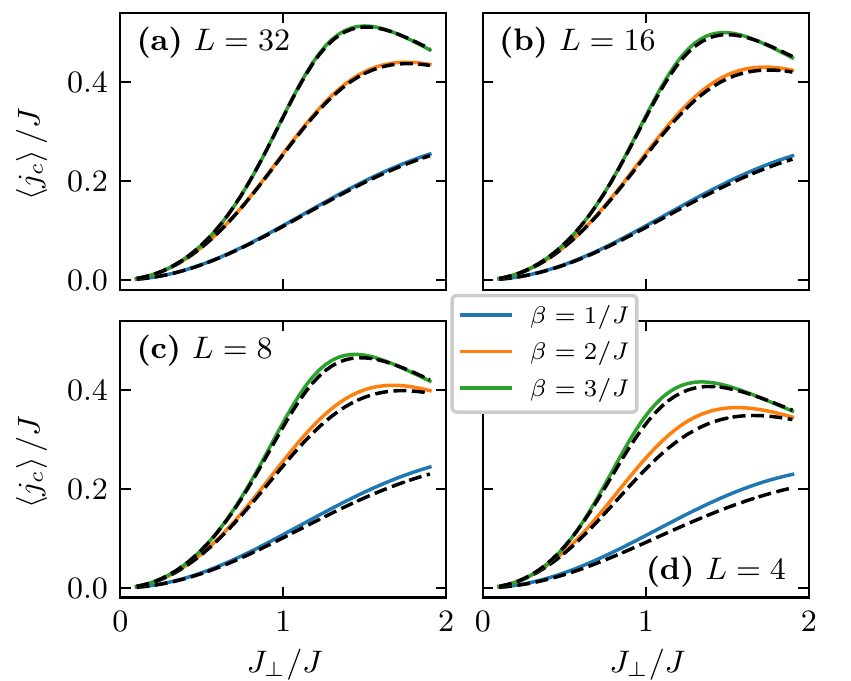}
	\caption{
		{\it Hard-core bosons: comparison of canonical and grand-canonical ensemble.}
		Chiral edge currents computed in the canonical (solid colored lines) and grand-canonical thermal states (black dashed lines) of hard-core bosons.
		We show data for $f=1/2$ and (a) $L=32$, (b) $L=16$, (c) $L=8$, and (d) $L=4$.
		In each panel, the lines correspond to inverse temperatures $\beta=1/J$, $\beta=2/J$ and $\beta=3/J$ (bottom to top).
	}
	\label{chiral_currents_can_gc}
\end{figure}

%
In Fig.~\ref{chiral_currents_can_gc}, we show that, for small ladder systems, chiral edge currents corresponding to canonical setups with fixed filling $f=1/2$ quantitatively differ from those corresponding to grand-canonical setups with average filling one-half.
On the other hand, for large systems, canonical and grand-canonical setups are expected to feature the same characteristics and the respective chiral edge currents should coincide. 
It can be seen in Fig.~\ref{chiral_currents_can_gc}(a) that for a ladder of length $L=32$ and considered temperatures $J$, $J/2$ as well as $J/3$, canonical (solid colored lines) and grand-canonical chiral edge currents (dashed black lines) overlap on the scale of the figure.
For small systems ($L=4$, $L=8$), chiral edge currents computed from the  grand-canonical systems are systematically smaller  than those computed  in the canonical ensemble.
In the following, we show only results obtained in the canonical ensemble.
\subsubsection{Integrated leg-gauge momentum-distribution functions}
\begin{figure}
	\includegraphics[]{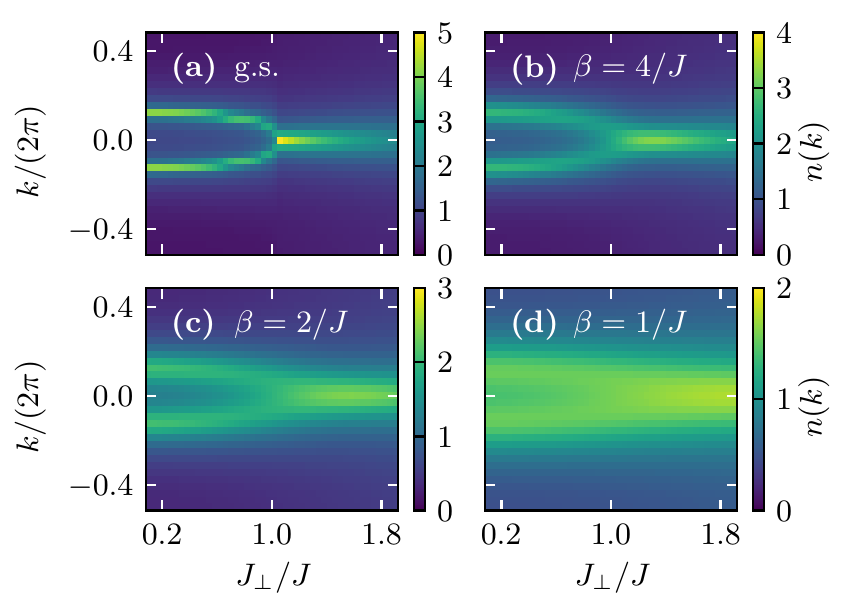}
	\caption{
		{\it Hard-core bosons, $f=1/2$, $n(k)$.}
		 Integrated leg-gauge momentum-distribution functions $ n(k)$ as a function of interleg coupling strength $J_\perp$ for  $L=32$ and open boundary conditions.
		(a) Ground-state (g.s.) momentum-distribution function at filling $f=1/2$.
		(b)--(d) Momentum-distribution functions corresponding to canonical thermal states at $f=1/2$ and different inverse temperatures (b) $\beta=4/J$, (c) $\beta=2/J$, and (d) $\beta=1/J$.
	}
	\label{mom_dists}
\end{figure}

Next, we consider integrated leg-gauge MDFs $n(k)$.
In this quantity, the ground-state vortex-fluid-to-Meissner transition manifests itself in the transition from two distinct maxima at finite momenta to a single maximum at zero momentum as shown in Fig.~\ref{mom_dists}(a).
There, we show results for  the integrated leg-gauge MDF computed in the ground state of a hard-core boson ladder with $L=32$ at $f=1/2$.
Figure~\ref{mom_dists}(b) demonstrates that the characteristic ground-state peaks persist in canonical systems with a finite temperature $J/4$ and that the two peaks in the vortex-fluid phase can be clearly resolved.
However, as expected, these sharp ground-state features blur out with increasing temperatures; see Fig.~\ref{mom_dists}(c) and Fig.~\ref{mom_dists}(d).
\subsubsection{Signatures of the vortex-fluid phase at finite temperatures}
\begin{figure}
	\includegraphics[]{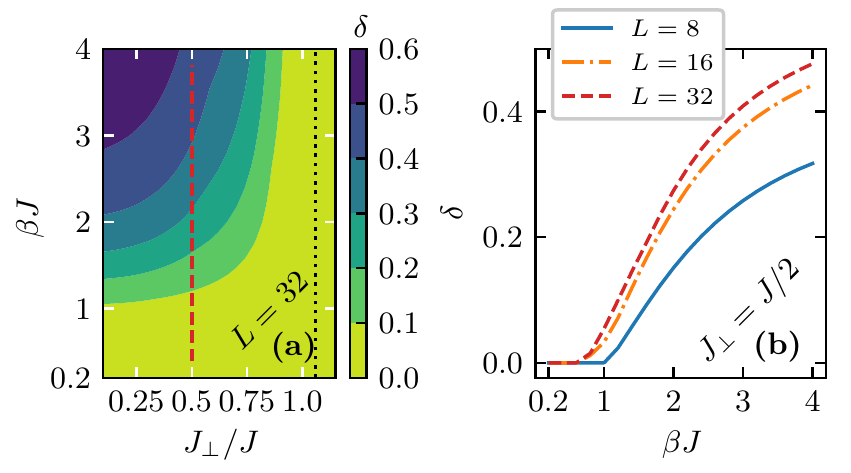}
	\caption{
		{\it  Signatures of the vortex-fluid  phase of hard-core bosons at $T>0$.} 
		(a) Contour plot of $\delta$, as defined in Eq.~(\ref{delta}), for different interleg coupling strengths $J_\perp$ and inverse temperatures $\beta$ for  a ladder of length $L=32$ and canonical thermal states at filling $f=1/2$.
		The value of the critical $J_\perp $ at zero temperature  \cite{piraud_2015} is indicated by a dotted line in (a).
		(b) Illustration of  finite-size effects: contrast $\delta$ versus $\beta J$ for ladders of length $L=8$, $16$ and $32$. 
		Panel (b) shows  a cut through  (a) at $J_\perp=J/2$ [dashed line in (a)]. 
	}
	\label{delta_hcb}
\end{figure}

Here, we concentrate on the measure of contrast introduced in Eq.~\eqref{delta} in order to quantify at which temperatures clear signatures of the vortex-fluid phase can be detected.
Considering a canonical half-filled hard-core boson ladder with $L=32$ rungs, Fig.~\ref{delta_hcb}(a) shows a contour plot of $\delta$ as a function of interleg coupling strength $J_\perp$ and inverse temperature $\beta$.
For temperatures below around $T\lesssim J$ and sufficiently small values of $J_\perp \lesssim 0.8J$, one finds $\delta > 1/10$.
This means that two different peaks, and thus a clear signature of the underlying vortex-fluid phase, can be detected. 
From Fig.~\ref{delta_hcb}(a), it also becomes apparent that, in the vortex-fluid regime, the contrast $\delta$ increases with decreasing interleg coupling strength $J_\perp$ and decreasing temperatures $\beta^{-1}$.
%

%
Moreover, the results for $\delta$ shown in Fig.~\ref{delta_hcb}(a) do not suffer from significant finite-size effects as can be seen in Fig.~\ref{delta_hcb}(b).
For the vertical cut at $J_\perp=J/2$ indicated in Fig.~\ref{delta_hcb}(a) (dashed line), Fig.~\ref{delta_hcb}(b) shows the contrast $\delta$ as a function of inverse temperature $\beta$ for ladders of length $L=8$, $L=16$ and $L=32$.
Note how sharply the curves for $\delta$ drop to zero as signatures of the two maxima in $n(k)$ are fully blurred out. 
Thus the derivative of $\delta$ with respect to $\beta$ would serve as a good indicator for the vortex-fluid-to-Meissner phase crossover.
A comparison of the $L=32$ data from Fig.~\ref{delta_hcb}(b) and Fig.~\ref{delta_nonint_bosons_L_128}(c) reveals that, for noninteracting systems, a finite contrast persists at smaller temperatures.
%
%
%
\section{Summary}
\label{summary}
In this paper, we investigated the properties of strongly interacting bosons as well as noninteracting bosons and spinless fermions which are confined on a two-leg ladder lattice subjected to a uniform magnetic field at zero and finite temperatures.
A particular focus was on the vortex-fluid-to-Meissner quantum-phase transition and the corresponding crossover observed at finite temperatures.
%

%
The chiral edge currents and momentum-distribution functions were the key observables analyzed in this paper.
In order to introduce the physics, the principal features and the analytic solution of a noninteracting ladder model with periodic boundary conditions were recapitulated \cite{paredes_2014,georges_2014}.
Our main results concern the vortex-fluid-to-Meissner crossover of hard-core bosons.
They were obtained by means of a matrix-product-state based purification approach following \cite{barthel2016matrix}.
This method is applicable to both canonical setups with a fixed number of particles as well as grand-canonical setups.
Typical chiral currents and momentum-distribution functions computed at zero and finite temperatures were discussed in detail.
%

%
We presented a comparison between results for  chiral edge currents obtained in the canonical ensemble compared to  those found in grand-canonical setups.
As a result, even for the relatively small particle numbers typical for quantum gases in optical lattices, (the technically simpler) grand-canonical simulations give quantitatively reliable results 
for the quantities of interest here and at sufficiently high temperature.
%

%
Moreover, we showed that clearly detectable signatures of the  underlying bosonic vortex-fluid phase persist in suitably chosen momentum-distribution functions at finite temperatures.
For this, a measure of contrast was introduced in Eq.~(\ref{delta}) that is sensitive to the presence of finite-momentum peaks in the integrated leg-gauge momentum-distribution function.
%

%
Considering the notorious difficulty of cooling a quantum gas to low-energy densities~\cite{McKay2011}, we expect our results to provide relevant guidance for future experiments which are naturally at nonzero temperature but allow control over the particle numbers.
Our results for the contrast can be compared to the resolution of a given experiment and will thus provide an upper limit for the temperature at  which the vortex-fluid phase is visible in the strongly interacting, low-density regime.
The methods and results presented here will also be relevant in characterizing the final state in quantum-quench dynamics and state-preparation protocols for the phases on flux ladders~\cite{langen14,polkovnikov11,eisert14}.
%

%
Our work paves the way for future investigations of finite-temperature properties of flux-ladder systems.
Open questions concern the finite-temperature properties of the bosonic ladder model in the presence of finite interaction strengths $0<U/J<\infty$.
In this regime, the model supports a variety of zero-temperature phases, including vortex-lattice and biased-ladder phases~\cite{wei_mueller_2014,greschner_2015_prl,greschner_2016}.
It has not yet been investigated if the characteristic signatures of these phases, such as typical current patterns and leg-population imbalances, persist at finite temperatures in the low-density regime.
The same applies to predictions for fingerprints of candidates for fractional quantum Hall states in these ladder systems~\cite{grusdt2014,petrescu_lehur_2015,cornfeld_2015,petrescu_piraud_lehur_2017}.
%
%

%
We thank M.~Aidelsburger, C.~Schweizer, and B.~Wang for useful discussions.
We are indebted to C.~Schweizer for his comments on a previous version of the manuscript.
This work was supported by the Deutsche Forschungsgemeinschaft (DFG, German Research Foundation) under Project No.~277974659 via Research Unit FOR 2414 and under Germany's Excellence Strategy -- EXC-2111 -- No.~390814868.
We acknowledge funding through the ExQM graduate school.
%
%
%
%
\bibliography{bibliography_ladder_project}
%
%
%
%
\end{document}